\documentclass[astrosymb, twocolumn]{aastex631}

\usepackage[utf8]{inputenc}	% allow utf-8 input
\usepackage[T1]{fontenc}	% use 8-bit T1 fonts
\usepackage{float}			% Allows for figures within multicol
\usepackage{comment}
\usepackage{enumitem}
\usepackage{graphicx}
\newcommand{\New}[1]{{\bf\color{red}{#1}}}

\newcommand{\HI}{\hbox{\rmfamily H\,{\textsc i}}}
\newcommand{\kms}{\hbox{km\,s$^{-1}$}}
\newcommand{\askapsoft}{ASKAP{\sc soft}}
\graphicspath{{./}{Figures/}}

\begin{document}

\title[DINGO: Options for storage]{Deep Investigation of Neutral Gas Origins (DINGO): Options for the Processing and Storage of Radio Astronomy Data for robust Deep Spectral Line Imaging in the SKA-Era using uv-Grids}

\author[0000-0003-0510-4311]{Alexander Williamson}
\affiliation{International Centre for Radio Astronomy Research, University of Western Australia, Crawley, WA 6009, Australia}
\affiliation{Australian SKA Regional Centre (AusSRC), University of Western Australia, 35 Stirling Highway, Crawley, WA 6009, Australia}

\author[0000-0003-0392-3604]{Richard Dodson}
\affiliation{International Centre for Radio Astronomy Research, University of Western Australia, Crawley, WA 6009, Australia}
\correspondingauthor{Richard Dodson}
\email{richard.dodson@icrar.org}

\author[0000-0002-6154-7224]{Pascal J. Elahi}
\affiliation{Pawsey Supercomputing Research Centre, Kensington, WA 6151, Australia}

\author[0000-0002-3570-4142]{Qian Gong}
\affiliation{Oak Ridge National Laboratory, Oak Ridge, TN 37830, USA}

\author[0000-0001-8496-4306]{Jonghwan Rhee}
\affiliation{International Centre for Radio Astronomy Research, University of Western Australia, Crawley, WA 6009, Australia}
\affiliation{Australia Telescope National Facility, CSIRO Space \& Astronomy, P.O. Box 1130, Bentley, WA 6102, Australia} 

\author[0000-0002-2838-3010]{Martin Meyer}
\affiliation{International Centre for Radio Astronomy Research, University of Western Australia, Crawley, WA 6009, Australia}
\affiliation{ARC Centre of Excellence for All Sky Astrophysics in 3 Dimensions (ASTRO 3D), Australia}

\author[0000-0002-5611-9292]{Krist\'of Rozgonyi}
\affiliation{International Centre for Radio Astronomy Research, University of Western Australia, Crawley, WA 6009, Australia}
\affiliation{ARC Centre of Excellence for All Sky Astrophysics in 3 Dimensions (ASTRO 3D), Australia}

\author[0000-0002-1774-5653]{Andreas Wicenec}
\affiliation{International Centre for Radio Astronomy Research, University of Western Australia, Crawley, WA 6009, Australia}

\author[0000-0002-1905-9171]{Jieyang Chen}
\affiliation{University of Oregon, Eugene, USA}

\author[0000-0001-9647-542X]{Norbert Podhorszki}
\affiliation{Oak Ridge National Laboratory, Oak Ridge, TN 37830, USA}

\author[0000-0003-3559-5772]{Scott Klasky}
\affiliation{Oak Ridge National Laboratory, Oak Ridge, TN 37830, USA}

\author[0000-0002-1828-1969]{Daniel Mitchell}
\affiliation{Australia Telescope National Facility, CSIRO Space \& Astronomy, P.O. Box 76, Epping, NSW 1710, Australia}



\begin{abstract}
The next generation of radio astronomy telescopes are challenging existing data analysis paradigms, as they have an order of magnitude more antennas and larger bandwidth.
Foremost amongst these are deep spectral line surveys, because these have the largest number of epochs and spectral channels per dataset.
For example, the Deep Investigation of Neutral Gas Origins (DINGO) project on the  Australian Square Kilometre Array Pathfinder (ASKAP) aims to observe over {3,200 hours spread over hundreds of observing sessions, covering two tiles, two footprints and two frequency settings}. 
The two primary problems encountered when processing this data are the need for storage and that processing is primarily I/O limited. 
To address these issues, we have implemented a deep imaging pipeline based on the storage of an intermediate data product in the software \askapsoft{}, that of the uv-gridded data, and have
demonstrated lossy and lossless compression of this data on ASKAP, using \texttt{MGARD} and \texttt{ADIOS2} libraries. 
We find data compression ratios from a factor of 7 (lossless) up to 20 (using lossy compression with an absolute error bound of $10^{-4}$), and processing is {significantly} faster for lossless compression. We discuss the effectiveness of lossy \texttt{MGARD} compression and its adherence to the designated error bounds, the trade-off between these error bounds and the corresponding compression ratios, as well as the potential consequences of these I/O and storage improvements on the science quality of the data products. 
As lossless compression allows us to achieve the DINGO goals within the storage limitations for the project, this will be the option adopted. 
\end{abstract}
% \begin{keywords}
% methods: data analysis -- techniques: interferometric -- Astronomical instrumentation, methods and techniques -- galaxies: evolution -- radio lines: galaxies
% %\RD{update Cosmology/HI/BigData}
% \end{keywords}

\section{Introduction} \label{sec:intro}
\subsection{The Square Kilometre Array}
Radio Astronomy is undergoing a paradigm shift
with the planning for a number of next-generation instruments, such as the Square Kilometre Array \citep[SKA,][]{Dewdney:2009}, the next-generation Very Large Array \citep[ngVLA,][]{ngvla_reference_design} and the next-generation Event Horizon Telescope \citep[ngEHT,][]{Doeleman:2023}.
All of these provide an order of magnitude increase in bandwidth and a few orders of magnitude in collecting area (and sensitivity) relative to  current radio telescopes.
This enhancement will provide us with the opportunities to survey the radio sky in exquisite detail, detect the signal from the epoch of reionisation when the first stars were born and measure the spectral signal from millions of galaxies.
To grasp this potential massive improvement in our understanding, the radio astronomy community must manage and process unprecedented volumes of data.
We are focusing on the SKA, as Australia is a founding member of the collaboration. The SKA will be built in Australia for frequencies spanning 50 to 350~MHz \citep[SKA-Low,][]{Labate:2022} and in South Africa for frequencies from 350~MHz to 15~GHz \citep[SKA-Mid,][]{Swart:2022}.
Phase-1 will have about 500 40~m aperture array elements in SKA-Low and about 200 15~m parabolic dishes in SKA-Mid. Construction has commenced and preliminary observations will be made from 2025. 
The data rates out of the correlators will be about 1\,TB/s, which will need to be captured in a local buffer and then processed  on the day - as storage will be limited \citep[see discussions in][]{ska-overview}.
%12EB/year (6.8/6Tb/s per site (40EB/yr); 122PB/yr HPSO weighted, in McMullin 20). \RGD{Do we want correlator output/data products? I think former.}

% \begin{itemize}
%     \item future specification
%     \item rough timeline
%     \item data sizes/flow rate

\subsection{Radio Astronomy Data}
% 1. Section 1.2. I suggest adding a few sentences about interferometry in general. Otherwise, the term "visibility" comes out of the blue.
{These next-generation instruments are all interferometers, as this is the most cost effective way to build a high resolution instrument.
Multiple (significantly cheaper) small dishes can be distributed spatially, to provide the high spatial frequencies in the image.
Interferometry consists of the measurements of the coherence function, at a spatial separation. In astronomy these correspond to the correlation of two signals, taken by antennas separated by a `baseline' vector and, as the signal is travelling from astronomical distances, represents the Fourier term of the transform of the brightness distribution of the observable sky, at the spatial frequency given by the baseline \citep[e.g., Van Citter 1934, discussed in][]{TMSv3}. % \citep{vancitter/prin_optics}. %\citep{TMSv3}[Chpt. 15]. % Or
%Larger separations (i.e. longer baselines) represent higher `spatial frequencies' and provide higher image angular resolutions.
%The more complete the sampling the more reliable will be the reconstruction of the sky brightness distribution. 
These measurements of the spatial coherence over the interferometric array are the visibilities, which are the output of the correlator that connects all of the stations in the array. }

Radio data presents a unique challenge. Much of the data is noise as the signal from any spatially concentrated region of high radio emission is spread between all of the visibility samples. 
Furthermore, the radio sky is largely empty and flux from the sources (expressed as T$_{\rm sky}$, the measured intensity in Kelvin) is much weaker than the visibility signals, which are dominated by the receiver temperature, T$_{\rm rec}$.
%, see for example Fig.~1 in \citet{dodson_25}. In this figure, only a small fraction of pixels in the image contain emission from a galaxy, which appears as a spatially concentrated region of high radio emission. 
% Not all astronomical sources are spatially concentrated and with ever improving resolution, what was once a single source can be resolved into spatially extended, diffuse emission. 
% Moreover, some signals, such as the sought-after signal of reionisation from the first stars, will be distributed across the entire image and is hidden in the noise. 

%\par 
This data analysis challenge is combined with a data volume challenge. Radio astronomy data volumes from current generation telescopes are of PB-scale. This data is also often stored as a Measurement-Set \citep{kemball:2000}, a format in which visibility and single-dish data are stored to accommodate synthesis. Although this format has been historically very useful, it does not scale particularly well and often the science process requires non-optimal access, giving rise to additional I/O load. 
%
%\par 
This data challenge will only increase once next-generation telescopes become operational.
%and is the motivation for this study. 
%HERE ADD STACKING/CO-ADDINF
The SKA strategy for multi-epoch (i.e. ``deep'') projects is to combine fully processed `daily' images. The data-reduction for image formation allows these to be stored for every observing session and then co-added (or image-stacked) to achieve the final sensitivity. 
The downside of this approach is that the deconvolution in the imaging, which is to correct for the imperfect instrumental sampling, is only partially applied and the final product is a combination of `cleaned' and `uncleaned' emission.
The traditional processing operates simultaneously with the individual datasets co-added to a common grid, which is imaged, fully deconvolved at the final sensitivity and model subtracted in major cycles. 
This obviously both requires all the dataset to be available simultaneously, which is well beyond scope for SKA and for the Australian SKA Pathfinder (ASKAP). 
We are investigating an alternative strategy {that} allows for full deconvolution yet does not require the retention of all the raw datasets. 
In this case we store a `daily' grid which is the residual visibility grid formed in the final major cycle of daily imaging, along with the response function and extracted model for that observation. This approach we call grid stacking, as we will co-add the uv-grids and do a final (minor) cycle of deconvolution at the final sensitivity. 
In principle, the uv-grids are not significantly larger than the image; there are the same number of voxels, with the former formed from complex numbers and the later from real numbers. 
{In the companion paper \citep{Rhee_25} we present a detailed comparative cube analysis between these methods.}
However, one \textit{also} needs to store the Point Spread Function (PSF) and the {Pre-Conditioning Function} (PCF), which are also stored as complex numbers. 
{The PSF grid represents the size, location and weighting of the complex-valued convolutional kernels applied during the gridding process. Both are used in the final imaging to apply weighting to individual visibility cells. 
The PCF grid represents the baseline sampling of the system, as the Wiener filtering requires the count of the samples per cell without kernel weighting. 
In this case the imaginary axis is used as an independent storage location for the average kernel size for that cell.}
The sum of these products can approach the same data volume as the original dataset.
Thus, for uv-gridding to be useful we need to compress these three grids; fortuitously these grids are largely made up of zeros, as the baseline sampling in a radio interferometer is very sparse and they compress very efficiently. 
% 2. page 2, right column, line 16. I stumble across "many non-linear processes" and that sentence in general. Application of calibration is perfectly linear (although figuring out of calibration correction is not) and so is the w-term effect itself (and also it is less of the problem for shorter baselines). Although I agree in general that baseline-dependent averaging may introduce additional problems to the data.
%{\bf discuss this with Mitch}
%% We absolute arragree that averaging the data before apply a linear correction will not introduce errors; however the uv-cell kernel is non-linear, and included in the w-term correction. We are meerly stating that BDC should be better than BDA, as an aside whilst we focus on the goals of the paper. We have improved the sentence to make clear that ..... calibration is usually linear (as it is interpolated between frequency and time sampling points); I do not see it mentioned however.

An alternative solution to the data volume problem is to apply baseline dependent averaging (BDA) to the data before processing \citep[e.g.][]{bda_stefan}. This is effective because short baselines have longer (slower) response times to the changes in the sky structure (and the atmosphere), because of their physical length.
{However, this means the data is averaged before correcting for any non-linear processes in Radio Interferometric imaging, for example some of the kernel corrections.}
Our approach is to apply these correction kernels to the raw data and then average to the uv-grid (which is the ultimate limit of BDA) and store these uv-grids in compressed format rather than the original visibility data. This is the approach we have implemented in the software for ASKAP and here we report on the impact of both the lossy and lossless compression options.

\subsection{The Australian SKA Pathfinder} %/Yandasoft}
The ASKAP \citep{Johnston:2007, Hotan:2021} radio telescope is one of the SKA precursors and is opening up a new window for large extragalactic {\HI} surveys beyond the local Universe due to its wide spectral bandwidth and large instantaneous field-of-view (FoV). ASKAP consists of 36 dishes, each of diameter 12 m and equipped with phased array feeds (PAFs) forming multiple (up to 36) receiving beams electronically \citep{DeBoer:2009, Hampson:2012}. The baseline lengths of the full array are from 22~m to 6.3~km. The phased array feed technology allows ASKAP to have a large $30~{\rm deg}^{2}$ FoV \citep{Bunton:2010} and a wide bandwidth of $288~{\rm MHz}$ with a channel resolution of 18.5-0.58~{\rm kHz} in the observing frequency between 0.7 and 1.8~{\rm GHz}, which makes it an optimal survey instrument, enabling it to conduct both wide and deep surveys in a comparatively short period of time \citep{Hotan:2021}.

\subsection{Deep Investigation of Neutral Gas Origins (DINGO)} \label{subsec:dingo}
Deep Investigation of Neutral Gas Origins (DINGO) \citep{Meyer:2009, Rhee:2023} is an ASKAP deep {\HI} survey project aiming to provide a cosmologically representative dataset for {\HI} emission, enabling studies of the {\HI} gas content of galaxies over the past 4 billion years.
%out to distances of 5 billion light-years, due to the accelerated expansion of the universe. 
The sky coverage of the DINGO survey is wider than deep {\HI} surveys previously conducted and the ongoing deep {\HI} surveys being carried out with other telescopes such as the JVLA\footnote{The Karl G. Jansky Very Large Array}, MeerKAT\footnote{The Meer-Karoo Array Telescope} and FAST\footnote{Five-hundred-meter Aperture Spherical Telescope} (CHILES\footnote{The COSMOS {\HI} Large Extra-galactic Survey \citep{Fernandez:2013}}, LADUMA\footnote{Looking At the Distant Universe with the MeerKAT Array \citep{Holwerda:2012, Blyth:2016}}/MIGHTEE-HI\footnote{The {\HI} emission project of the MeerKAT International GigaHertz Tiered Extragalactic Exploration survey \citep{Maddox:2021}}, and FUDS\footnote{FAST Ultra Deep Survey \citep{Xi:2021}} respectively). Due to its large volume coverage (60 deg$^2$ compared to 20 deg${^2}$ for MIGHTEE \citep{Maddox:2021} with a similar sensitivity and redshift range), the DINGO survey will reduce cosmic variance on {\HI} measurements, thereby providing a unique legacy {\HI} dataset.

%\MM{\sout{I would reorganise the below paragraphs slightly - starting with the description of the full survey and data problem, then describing the 200h data used in this paper.}} \RD{Addressed?}

DINGO observations are focused on the Galaxy and Mass Assembly (GAMA) \citep{Driver:2022} 23~h (G23) field, centred at $\alpha, \delta$ (J2000) = $22^{\rm h}59^{\rm m} 00 \rm \fs00, -32\arcdeg18\arcmin00\farcs0$. These observations used the full array of ASKAP's 36 antennas with the 288~MHz bandwidth (15,552 channels) in the observing frequency ranges of 859.5-1147.5~{\rm MHz} (band 1) and 1151.5-1439.5~{\rm MHz} (band 2). 
% 5. page 3, left column, lines 23-24. Do you mean two telescope pointings A and B? For ASKAP, the term "footprint" is typically used for the beam arrangement (i.e.
The channel width is 18.5~{\rm kHz}, equivalent to a velocity resolution of $\sim 3.9(1+z)$~\kms in cosmologically nearby galaxies. 
The DINGO full survey allocation is 3,200~hr, split into two 1,600~hr allocations for the two frequency ranges of band 1 ($z \sim$ 0.2 to 0.4) and band 2 ($z$ of 0 to 0.08) and alternating between two tiles, each with {two dithering-footprints, A and B, stepped in a half beam-width,} to ensure more uniform sky sensitivity.
%These 
%\sout{aims to obtain} 
%100~hr of data are split between the band~1 and 2 frequencies on the G23 field, to develop the DINGO processing pipeline for deep imaging and long-term data storage. 
%
% From Rhee 25
We would expect the final data volume will be from 100\,days per tile, each of 8\,hours and with 36\,beams and 2\,bands.
As each MS file is 57\,GB/hour in size the storage requirement for the full dataset will be 6.6\,PB, which is beyond feasibility. 

The default ASKAP and SKA strategy to address this is to image each day and stack these reduced data products to reach the results. 
Numerous investigations have suggested that this approach would be sub-optimal \citep[for example,][]{Dingo_Memo4,dodson_16,Rozgonyi:2021}. This has now demonstrated in \citet{Rhee_25}, which is the companion data-analysis paper for this data-methods discussion, where the results from grid-stacking and image-stacking are compared to the traditional approach.
What \citet{Rhee_25} shows is that neither of the stacking approaches fully recover the results from traditional imaging, because the deconvolution in the imaging process is non-linear. This leads to a 1\% shortfall in the recovered flux in the grid-stacking compared to traditional, but a 8\% shortfall for image-stacking. The latter also suffers from significant image quality degradation (see Figures 5 and 3 in \citet{Rhee_25}, respectively).
The reason for this is that no further deconvolution is possible in image-stacking after the daily deconvolution, whereas in grid-stacking further minor-cycle deconvolution is possible but major cycles is not. 
This means that the weaker sources in the image-stacked cubes, in particular the low-column density HI, is a mixture of deconvolved and non-deconvolved emission, whereas this problem is much reduced in grid-stacking (and traditional) processing.
%for the latter it means that the final deconvolution would still suffer from any non-linear or kernel-based effects. 

{There are further approaches that could be developed further.} % in principle, 
One could perform a further minor cycle on the image-stacked residual cubes \citep[as discussed in ][]{Rhee_25}, but one would now need to store the daily residual image, the models and the image-domain PSF (at four times the image size). These would now be larger than the original data-products and would not be highly compressible as there are no longer empty cells.
The stacked uv-grids could be extended to support major cycles, by extending the dimensionality, either by separating the data by w-plane or by baseline. As an uncompressed product these would be much larger than the original data, but as the number of occupied cells would not increase significantly \citep{Dingo_Memo4} they should compress to more or less the same storage footprint.
Neither of these ideas have been explored in this work nor are currently supported by \askapsoft{}.
% 4. page 3, right column, lines 20-24. Although I agree that, strictly speaking, there are "no longer empty cells" in the image-domain products, one can always FFT and there should be holes (although perhaps filled a bit by non-linear deconvolution). It is possible that some compression algorithm is clever enough to do equal compression without the explicit FFT (as in principle, there is the same amount of information in both). Having said that, I agree with the main point that storing grids has an advantage here (although it may be easier to store the original data than to increase dimensionality storing grids per baseline or w-plane).

So far a total of $\sim$200~hr observations have been conducted in the higher frequency band, between 2019-2024 and with both footprints for one tile.
% 6. page 3, right column, line 34. I stumble across "25 individual MS". Do you mean individual observations (or scheduling blocks) here? ASKAP writes data split by beam and by 48 MHz of bandwidth, so even a single observation would produce 216 measurement sets.
These are formed of 25 individual {$\sim$8~hr long observations, consisting of} 36 beams and two footprints, with 15,552 channels spanning 288\,MHz, centred on 1295.5\,MHz, of which only the upper half (above the RFI from the GPS satellites) was processed.
See Table~1 in \citet{Rhee_25} for full details.
We are using these early observations to test the compression options for the proposed alternative strategy of uv-grid stacking. 
%xxxxxxxxx

\section{Methods} \label{sec:methods}
\subsection{ADaptive I/O System version 2 (ADIOS2)}
The Adaptable Input Output System version 2 \citep{Godoy:2020}, is a software framework with a simple input/output abstraction and a self-describing data model centred around distributed data arrays, allowing multiple applications to publish and subscribe data at large levels of concurrency. It also introduces a larger organizing concept, the ``step'', for driving data production and consumption within applications. 
%
% In particular, one obvious missing piece of information is the description of ADIOS2 image format, which is different from the standard CASA image
% in that it does not write everything into a single table cell. This is the fundamental difference enabling parallel I/O via ADIOS2 (and, in fact, it is
% possible to parallelise write by tiles in the ordinary tiled storage manager if the data format changed to an analogous multi-cell version). This
% directly affects section 3.2. The number of table cells used to distribute the cube data across is also a crucial parameter potentially affecting the performance.
%
{{\texttt{ADIOS2} is primarily focused on high-performance and parallel I/O, with its parallel storage performance, the file format, the memory management, and data aggregation algorithms being designed together to be highly scalable in every axis (many processes, many variables, large amounts of data, many output steps). 
Our I/O backend is implemented using the file-based storage I/O engine in the \texttt{ADIOS2} library. The engine designates separate MPI ranks as aggregators, which collect data in a streaming fashion, buffering them in large chunks before writing to a disk or a solid-state drive to avoid small I/O requests. 
As such the data can be independently written by each MPI imaging process, which in the DINGO analysis is many threads that each manage the data reduction for a small range of channels (18 in the analysis presented here). 
Thus \texttt{ADIOS2} provides a native parallel I/O library for high performance high speed streaming data transport
and the \texttt{ADIOS} image class distributes data between multiple table cells, allowing us to fully utilise the high-performance compute platform.
The self-describing data are stored in binary-packed (.bp) format to allow for rapid metadata extraction. 
Of particular relevance to the astronomical community is that it can be embedded in the \texttt{CASACORE} libraries as an alternative storage manager (if enabled at compile time), and is able to write tables for both images and visibilities. 
We note that the ADIOS Image Class is not, strictly, in \texttt{CASACORE}, but is `research code' currently available via the \texttt{CASAREST} library. A pre-compiled DINGO-ASKAPSoft image is available on request and we expect this functionality to be eventually available in the default ASKAPSoft release.
For the DINGO processing the raw input data column (typically ``DATA'') could be in an ADIOS format, but it is not in this pipeline; however the image cube is, and written out as columns ``map'' and ``mask''.
}}
\texttt{ADIOS2} recently developed a new mechanism to allow applications to use state-of-the-art lossless and lossy compression algorithms. This mechanism makes use of tight integration between I/O and reduction and allows applications to take full advantage of the self-describing formatting and lossy compression techniques.

\subsection{MultiGrid Adaptive Reduction of Data (MGARD)}
\texttt{MGARD} \citep{Gong:2023} offers error-controlled lossy compression rooted in multi-grid theories. It transforms floating-point scientific data into a multilevel representation, followed by quantisation, lossless encoding, and ultimately generating a self-describing compressed buffer. %The presence of random mantissa in floating-point representations presents challenges for lossless compressors when attempting to reduce scientific data. MGARD can efficiently reduce floating-point data and guarantee the compression incurred errors below user-prescribed error bound. 
One of \texttt{MGARD}'s notable features is its array of error control options, including $L^\infty$, $L^2$, point-wise relative $L^\infty$, and options to define varied error bounds across regions or different frequency components. This flexibility is valuable for preserving Quantities-of-Interest (QoI) \citep{Gong:2022} derived from the reconstructed data. For region-adaptive compression, \texttt{MGARD} accommodates Regions-of-Interest (RoI) specified through either bounding boxes or masks, with the latter especially useful for irregular shaped RoIs. In cases where RoI information is not provided, \texttt{MGARD} employs internal functions to identify regions rich in detail, leveraging data turbulence measured across multiple scales.       

\subsection{\askapsoft}
\askapsoft{} is a package that contains the software necessary for processing data from the ASKAP telescope. Its primary purpose is for the full-scale processing of ASKAP data, from the observed visibility data to spectral-line and continuum images.
\askapsoft{} provides the well-established routines to image a spectral line dataset. That is to: {\it read} the data, apply {\it weighting} kernels to set the image parameters (Field of View, sensitivity to low surface brightness or compact objects, etc), resample the data onto a regular sampled {\it grid} for {\it inversion} using the Fast Fourier Transform, iteratively {\it deconvolve} for the limited sampling of Fourier terms and finally, for deep images, {\it stacking} of multiple epochs of observing for the final image.
The DINGO pipeline reorganises these tasks so that grids are preserved for stacking, rather than the images. This reduces the number of inversions required and improves the quality of the deconvolution.

Imaging radio interferometric data at SKA-scales is expected to be I/O bound due to the massive size of the datasets. The computational costs are dominated by the gridding and the inversion steps, and these two are expected to have approximately similar requirements. Thus any reduction in the size of the datasets would have a significant impact on the total processing time and any reduction in the gridding or inversion would have a significant (albeit smaller) impact on the compute costs.

\askapsoft{} contains a wide variety of programs and scripts that are useful in the analysis, manipulation, and processing of radio astronomy data. We will primarily use two applications present in \askapsoft{}: \texttt{imager} \& \texttt{cdeconvolver}. \texttt{imager} creates spectral-line image cubes, which use frequency as an analogue for distance, allowing for a 3-dimensional view of the sky. 
Thus this single program includes the reading, weighting, gridding, {deconvolution} and inversion steps.
%
%{In the companion paper \citep{Rhee_25} we present a detailed comparative cube analysis between our methods discussed here (`grid stacking') and the `traditional' and the `default' ASKAP methods, those being: the approach of processing all visibilities together to produce a single final image and imaging one \New{epoch} of visibilities at a time to produce a daily image that are then averaged together (`image stacking').
{For the investigations here we used identical imaging parameters parameters as in \citet{Rhee_25} for \texttt{imager}: 30\farcs resolution and 6\farcs cell size, 3 major cycles each terminating when the 5$\sigma$ noise-level is reached, except for regions where flux has already been cleaned, in which case the threshold is 0.5$\sigma$.}
In our case, \texttt{imager} is used to produce visibility grids from each schedule block, as an intermediate product, which we then combine to produce the final image. 
That is, the {final} inversion step is not performed and the normally intermediate data products (i.e. the residual grids and the model components subtracted) are saved for later processing.
Due to the sparse nature of these grids, compression is particularly efficient. 
Furthermore, in principle, at this point the image
%and is at a sufficiently early point in the pipeline, so that 
processing parameters can be changed as needed to adjust weights and even field of view. The deep imaging mentioned in section \ref{subsec:dingo} requires stacking these grids over 3,200 hours of observed data to improve sensitivity to sources within the final image.
For this task we use
\texttt{cdeconvolver}, a bespoke application designed specifically to perform this stacking and complete the imaging process. Further information on these applications can be found at \textcolor{blue}{\url{https://www.atnf.csiro.au/computing/software/askapsoft/sdp/docs/current/index.html}}

\subsection{DINGO Pipeline}
\begin{figure}
    \centering
    \includegraphics[width=\linewidth, trim=1.cm 10.25cm 2.cm 2.cm, clip=true]{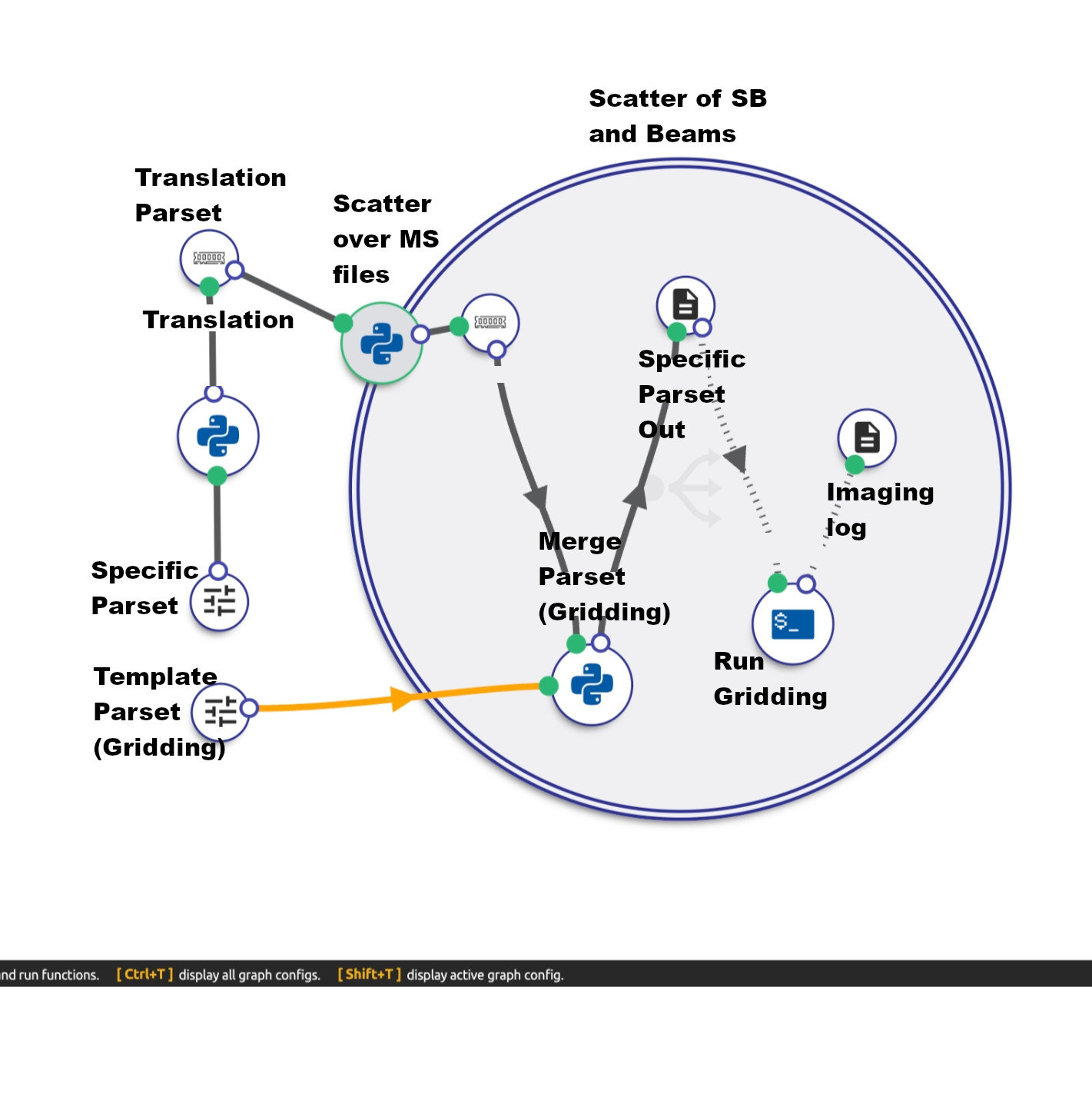}
    %width=\linewidth, trim=1.cm 1.25cm 2.cm 2.cm, clip=true]
    \caption{{The graphical representation in EAGLE of the DINGO gridding pipeline. Specific parameters are provided, split among the scattered instances, merged with the global template and the individual \askapsoft{} imager applications are run to generate the schedule block residual visibility grids and the model components subtracted. Following this, in another similar graph, the gridded visibilities are combined, imaged and the model components are added back to produce the final stacked image.}}
    \label{fig:eagle}
\end{figure}

%\AJW{has agreed to revise.
%Expand this section to become a reference point.
%DALuiGE etcetc. EG:
{The DINGO workflow is designed using the DALiuGE framework \citep{daliuge}, which is a scientific workflow development and execution framework to process large astronomical data sets at a scale required by the Square Kilometre Array, with a particular focus on efficient data I/O management. 
The executed workflow is represented by a Directed Acyclic Graph (DAG), where the vertices are an alternating sequence of stateless computational tasks and data nodes, edges connect the output of one task with one input of a data node which in turn connects to the input of the subsequent task. 
The graphs can be developed using the graphical editor EAGLE\footnote{http://eagle.icrar.org}, allowing the intuitive representation of complex data reduction workflows, consisting of both datasets and algorithmic components. EAGLE also allows for the run-time implementation and monitoring of the execution of the pipelines on distributed resources. 
By mapping the logical view of a pipeline to its physical realisation, DALiuGE separates the concerns of multiple stakeholders, allowing them to collectively optimise large-scale data processing solutions in a coherent manner.
Figure \ref{fig:eagle} shows the pipeline as constructed in the EAGLE GUI for DALiuGE, where we built the pipeline from components autoloaded from the codebase (using, for example, Doxygen/Sphinx formatted component documentation for the derivation of the input and output parameters). 
}

% 7. page 5. left column, line 53. Do you really mean "independent of frequency"? Standard ASKAP data processing takes a couple of Taylor terms into account, i.e. not just Taylor term 0.
We start with a MeasurementSet data format that contains the calibrated, continuum-subtracted visibilities (the continuum here refers to the components of the data that are {not limited in frequency span}). This is the input to \texttt{imager}, described above, which produces a residual visibility grid, a PSF grid, a PCF grid and the spectral model components subtracted to produce the residual visibilities. The visibility grid is a grid representation of the 3-dimensional visibilities projected onto a 2-dimensional grid for each frequency channel (producing a 3-dimensional grid). 
%The PSF grid represents the inherent convolution of point sources due to the baseline sampling of the system. The PCF grid represents the size, location and weighting of the convolutional kernels applied during the gridding process. These are used in the final imaging to apply weighting to individual visibility cells.
These grids are passed to the \texttt{cdeconvolver} application which, if more than one observation is provided, will sum the grids together as they are read in. These summed grids are then imaged with minor cycles for analysis. 
%Here we validate that the resulting image is free of detrimental RFI and that the sensitivity of the image is not degraded.
%better than that of the non-stacked image.
\subsubsection{Performance Measurements}

{In \citet[Table 3]{Rhee_25} we have made detailed comparisons of the scientific outcomes of grid-stacking, in comparison to both traditional processing of all the daily epochs simultaneously, i.e. without consideration of storage limits, and the alternative storage-constrained approach of image-stacking.
The focus of this paper is the performance of ADIOS2 and lossless and lossy gridding, which obviously only is relevant for grid-stacking.
%But for the comparison between the science outcomes of the different approaches we have measured the other major performance metrics for all three approaches: memory footprint, I/O and compute time. 
%We summarise these in Table \ref{tab:results-full_resources}. 
The use or otherwise of ADIOS has minimal impact on the required Memory, compute and read I/O, which are dominated by the management of the raw visibilities. 
For the processing of a daily image-cube with grid stacking we found that processing 7776 channels required 80min on 217 cores, with a maximum memory footprint of 30GB, with a required read I/O of 40GB. 
The write I/O without using ADIOS was about 30GB. The fractional impact of ADIOS on the latter figure is investigated in Section \ref{sec:results}.
In addition to the spectral line imaging step measured here, the grid-stacking requires the models to be restored, and all methods require subsequent continuum subtraction in the image domain to remove residual continuum emission, and two stages of linear mosaicking,
but these are not significant additional computational cost.}

\subsection{Compression}
% 8. page 5, section 2.3. Perhaps, it is worth stating up front that deconvolution job may be more picky on the PSF quality. The rest of the section kind of supports that this is the reason behind issues with the 10^{-3} error bound, but it is only stated explicitly in section 3.1. I also wonder if it would be more forgiving in terms of the high error bound, if one Fourier transforms the PSF before the compression (due to reduced dynamic range requirement). By the way, I do not see huge reason to discuss both absolute and relative error bounds here and later in the paper. Perhaps, just state that absolute bounds behave as expected given the range of the data.
%%%%%%%%%% agreed
The focus of the compression here is that of the grids. Due to the limited range of the values in the PCF grid (i.e. limited to half integers), lossless compression provided a compression ratio of $\sim$100, with lossy compression providing ratios similar to that of the visibility and PSF grids. For this reason, we are only comparing the compression of visibility and PSF grids.
We use lossless and lossy compression with error bounds of $10^{-3}$, $10^{-4}$, $10^{-5}$, and $10^{-6}$ both as relative and absolute error bounds, {although these metrics are somewhat overlapping.} % as the relative error bound being an absolute error bound scaled by the data range (defined as half the maximum value less minimum value).
The \texttt{cdeconvolver} application failed to complete the imaging with a relative error of $10^{-3}$, due to a failure to converge. 
{We believe that this is due to the errors accumulating in the PSF reconstruction which, as it is applied iteratively, is most sensitive to numerical error.
To investigate this further we would have used separate compression scales for the PSF as opposed to the visibilities,  but this was out of scope of our investigations.
}
%, the reasons for which are still under investigation.
%
The relative error bound is derived against the range of the data being compressed (i.e. against half of the minimum to maximum data values) whereas the absolute error bound is applied without scaling.
The typical range of values for the input datafiles are about $\pm$100, so the absolute error bounds are two orders of magnitude less than the relative values.
For the application to gridded data we quickly discovered that it was vital to ensure that zero values were not changed to values near zero but within the error bound, which is the default behaviour of multi-grid methods. 
This was solved by compressing both the raw data and a mask which records the position of the zeros.
%\QG{This problem can be solved by compressing both the raw data and a mask which records the position of the zeros. I previously verified that the size of the mask was tiny (e.g., over 100x of reduction) after compression.} 
This is vital because the sum of visibilities in a grid cell is normalised against the weight of samples included in that sum, as stored in the PSF/PCF grid. Small near-zero values causes the imaging to diverge. 
%\QG{Regarding MGARD compression results shown in this paper, did we apply a mask to ensure the zero values in the original data to remain as zeros after lossy compression? }
%
The lossless compression uses the zstd algorithm to compress the grids, although bzip2 also provides a similar level of compression. This provided a consistent compression ratio and (as expected) did not alter the decompressed data in any way.

\subsection{Parallel I/O}
The integration of \texttt{ADIOS2} into the I/O stage of \askapsoft{} leveraged the use of \texttt{CASACORE} and the implementation of the existing \texttt{ADIOS2} storage manager \citep{WANG:2016}. An \texttt{ADIOS2} Image module was developed to bridge the gap between \texttt{imager} and \texttt{cdeconvolver}'s use of image inputs and outputs, and the storage manager's table interface. 
These applications use MPI to implement parallel processing which, when passing the communicators to the storage manager (and by extension \texttt{ADIOS2}), enables the I/O to occur in parallel.
%Not true!
%The impact on \texttt{imager} was small, as the main I/O is the raw data that continues to be in standard format, as opposed to \texttt{cdeconvolver} where the input format could be standard, adios-single threaded or adios-parallel threaded.
{These capacities have been or are in the process of being merged back into the \askapsoft{} package.}

\section{Results} \label{sec:results}
\subsection{Compression Comparison}
\begin{figure}
    \centering
    \includegraphics[width=\linewidth, trim=1.cm 1.25cm 2.cm 2.cm, clip=true]{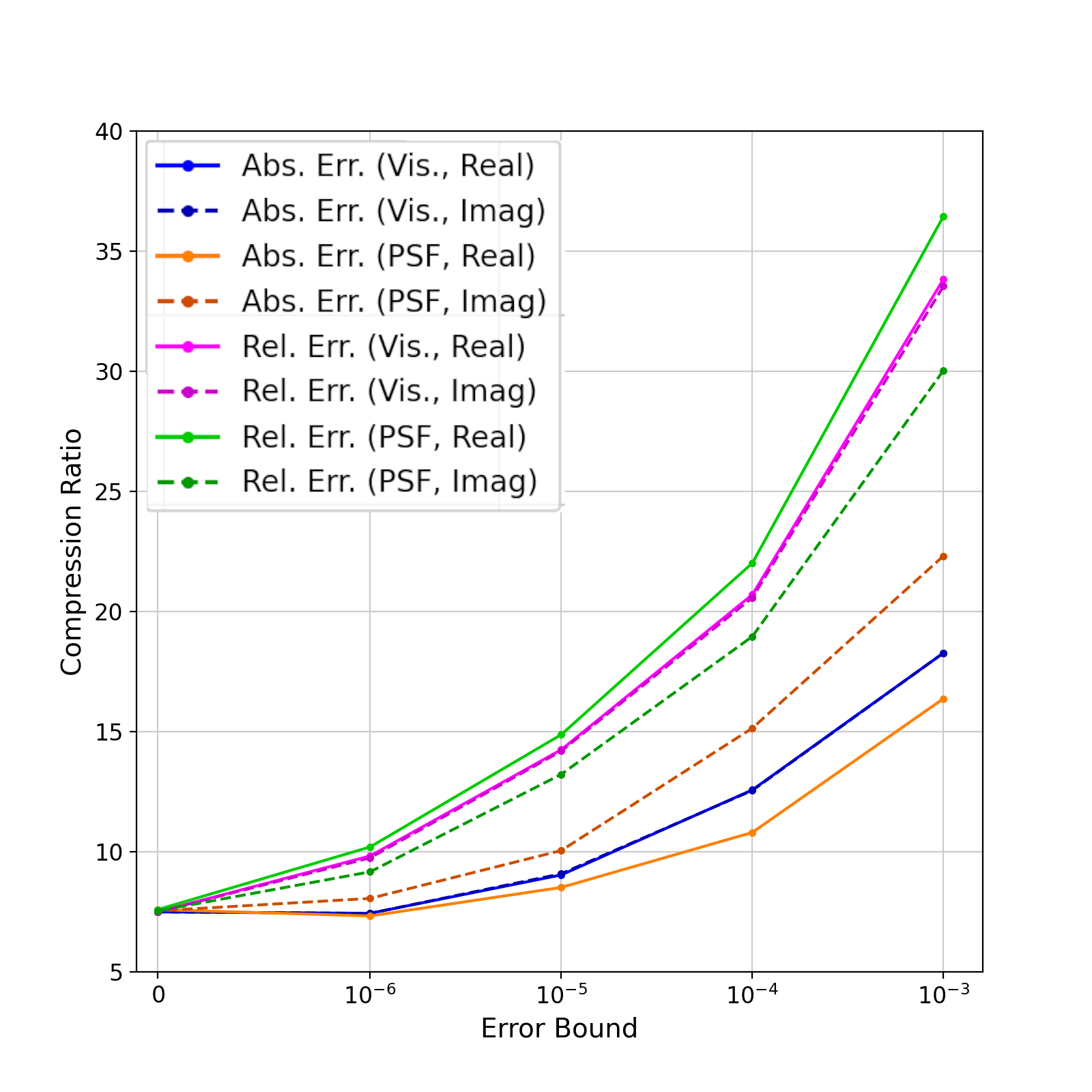}
    \caption{The compression ratio for increasing specified error bounds. Shown are the visibility and PSF grid compression ratios for lossless compression, lossy compression with absolute error bounds, and that with relative error bounds.}
    \label{fig:eb_cr}
\end{figure}

% 9. page 6, section 3.1. Is it possible to explain the lossless compression factor through the number of empty cells in the uv-coverage?
%%%%%%%%%% agreed, we studied this in simulations. Plot attached
Figure \ref{fig:eb_cr} shows the compression ratios of the visibility and PSF grids as the error bound are increased from zero (lossless) to $10^{-3}$ for both absolute and relative error measurements. 
{Lossless compression is consistent at a value of 7.5, whereas the lossy compression ratio increases from that limit with increasing error bound,} and the compression ratios of the real and imaginary parts of the visibility grid are consistent. 
The absolute error bounds for the visibilities are effectively more than an order finer than the numerically equivalent relative error bound, and this can be seen in higher compression for the relative results. 
%\New{The impact of the fraction of the {\it uv}-grid that was empty on the compression ratio would be close to linear with ADIOS, as a mask for zero-valued cells is formed and separately compressed. This was to ensure that unobserved {\it uv}-grid points never contributed to the  {\it uv}-coverage (and alter the image) through the compression approximation.} 
%whereas the imaginary parts compress better than the real part in the absolute case and worse in the relative case. 
%\QG{I think the statement of ``better than the real part in the absolute case and worse in the relative case'' was misleading. The reason why the compression ratios were high for imaginary parts than real parts under the same absolute error bound is because the value ranges are different for real and imaginary data. I think we should only use relative error bounds to be consistent, unless the same percentage of errors in real and imaginary parts have disproportionate impact on the final image.} 
%This is due to the fact that the real and imaginary parts of the PSF (unlike the visibilities) represent different properties and cover a different range of values.

Of the eight tests of the compression, seven completed the imaging and produced, qualitatively, decent images. The reason for the failure of relative-$10^{-3}$ result is %still under investigation, however this is 
{likely caused by the numerical degradation of the PSF, and thus errors introduced in the convolution, and a divergence in the image deconvolution.}
%ue to aa mismatch between the deviation in values between the resulting grids, after compression. 
%These images are shown in Figure \ref{fig:imgsum}, which shows a collapsed view over the 60 channels that contain a radio source. The only images in this panel that show any qualitative deviation from the original (or lossless) image is that for the $10^{-4}$ relative error bound and the $10^{-3}$ absolute error bound.
% \begin{figure}
%     \centering
%     \includegraphics[width=\linewidth, trim=0.cm 2.0cm 0.cm 2.cm, clip=true]{images_vertical.png}
%     \caption{Images produced by \texttt{cdeconvolver} from visibility grids compressed with \texttt{MGARD} at different error bounds. The lossless image (top right) is identical to the original and can be used as a ground truth in this case. Bright emission is dominated by single source in this example image.}
%     \label{fig:imgsum}
% \end{figure}

\begin{figure*}
    \centering
    \includegraphics[width=\linewidth, trim=4.cm 0.5cm 4.cm 0.5cm, clip=true]{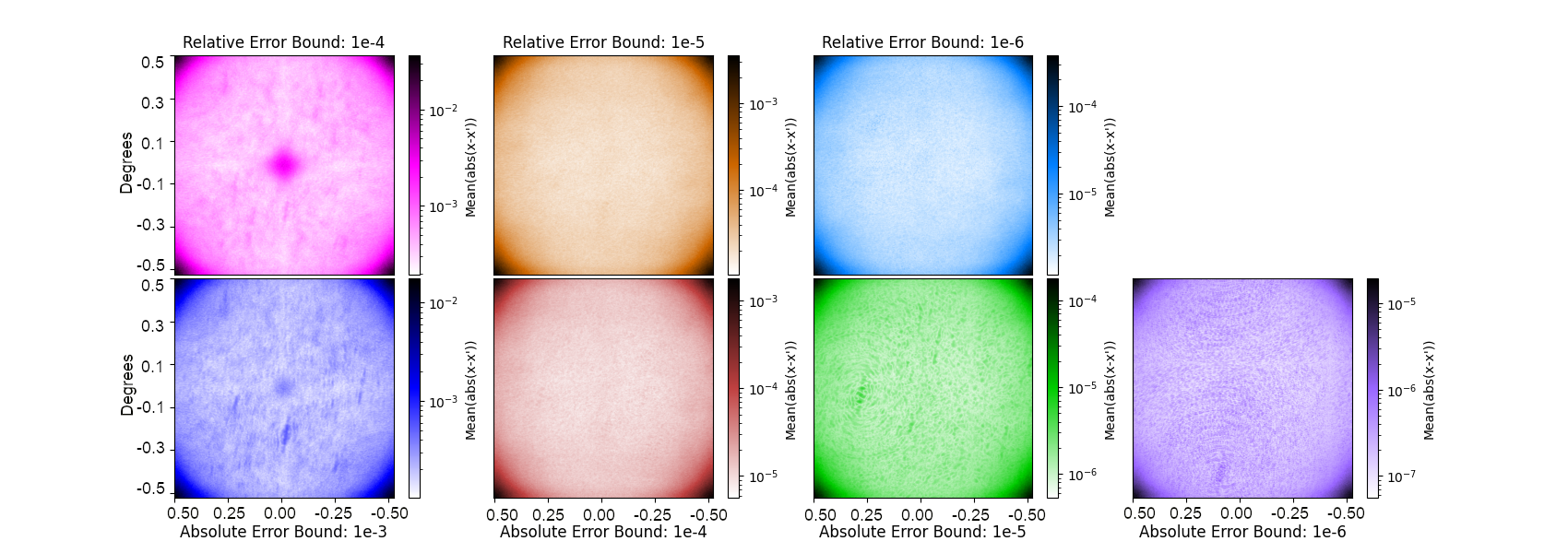}
    \caption{The absolute residuals between the images produced after lossy compression and the uncompressed equivalent. The colour scales are not normalised as the range of values for each image is significantly different. Ignoring the image corners, the
    smallest error bounds on the right-hand side lead to maximum residuals of $\sim10^{-5}$.}
    \label{fig:imgdiff}
\end{figure*}

%\sout{\RD{Old Fig 3 removed following discussions with MM}}
The images produced from the compressed data are visually close to identical; it is only in the image differences that the impact of compression can be seen.
%This is further reinforced by 
Figure \ref{fig:imgdiff}, plotting the difference between the images made with lossy compression and those without,  shows that majority of the value deviation is at the corners of the image. 
Correcting for primary beam weighting across the FoV would exclude these points. {We suspect this problem has a common origin with the complete failure of the image-convergence at relative error bounds greater that $10^{-3}$.}

%\QG{Was the image showing the frequency domain and the corners represent high frequency channels?} 
For the more aggressive compression, i.e. the $10^{-4}$ relative and $10^{-3}$ absolute error bounds, increased residuals near the centre of the image appear, which would have an impact on the science outcomes. At higher precision error bounds the errors introduced across image are uniform.
%For a detailed study of the impacts of error bounds into the recovered image properties see \citet{dodson_25}. 
% O 11. page 6, right column, lines 33-37. It would be better to refer to physical units (e.g. in wavelengths) instead of the pixel numbers (and the same for plots). When discuss the 30 arcsec resolution, perhaps mention that DINGO (and other spectral line ASKAP projects) currently only offered 2km baselines in the standard processing (otherwise some readers may expect finer resolution based on 6km maximum baseline of the instrument).
%%% Whilst I absolutely agree, this is one of the challenges. We have only the results as left by Alex when he left. Thus I have editted the PNG to get this plot. 

\begin{figure}
    \centering
    \includegraphics[width=\linewidth, trim=1.cm 1.cm 2.0cm 3.cm, clip=true]{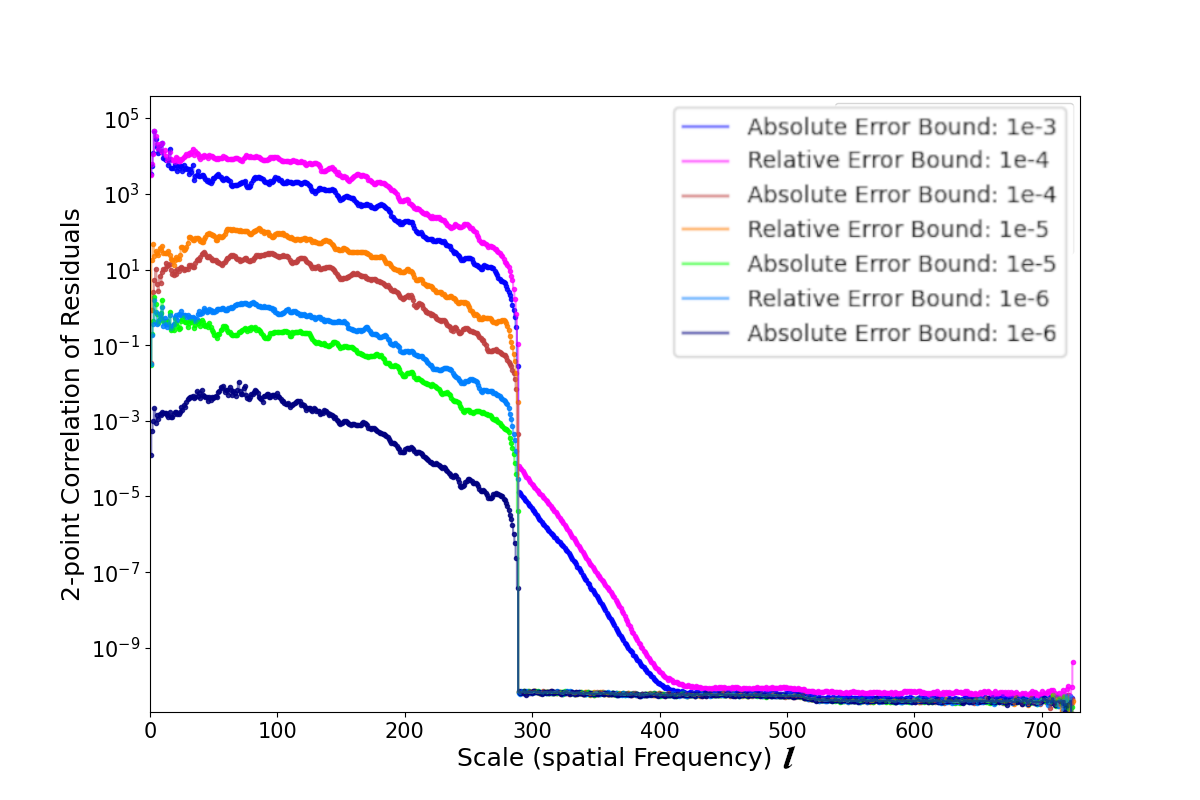}
    \caption{The unnormalised 2-point correlation of the residuals of the images produced with varying error bounds. The x-axis represents the spatial frequency bins of the image, {where $\ell$=0 is equivalent to structures over the whole field of view and $\ell$=700 is equivalent to structures of 12\arcsec, double the pixel size.} The colours are as for Figure \ref{fig:imgdiff} and same first two error bounds show power at angular scales greater than the resolution, indicating reconstruction errors.}
    \label{fig:2pcorr}
\end{figure}

% 10. page 6, right column, around line 30. Perhaps, "square of the absolute value" would be better than the "product of the value of these pixels with their conjugate". How were the values in the given annulus combined? Before or after square of the norm was calculated?
An alternate distortion measure is the 2-point correlation of the residuals. This can be produced by performing a Discrete Fourier Transform (DFT) on the residuals, {coherently binning the pixels in radius and calculating the square of the absolute value of these pixels. This is mainly to detect ripples in the residuals from poor deconvolution and/or calibration errors.} The result is shown in Figure \ref{fig:2pcorr} and describes the prominence of patterns of certain scales within the residuals. The higher error bounds all show a sharp cutoff at {$\ell$=289, which corresponds to the largest radial distance of a non-zero cell on the visibility grid, that is equivalent to the resolution -- 30\arcsec}, whereas the residuals for highly compressed data with $10^{-3}$ absolute error and $10^{-4}$ relative error show a negative logarithmic trend between $\ell$ 289 and 410. 
%The value of 289  and corresponds to the smallest resolvable scale, i.e. 30\arcsec. 
An excess of values above this spatial frequency shows a leakage of values within the PSF grid, adding small scale residuals below the smallest resolvable scale, {and is found only for PSFs which have been over-compressed}.

% The spectral profile shown in the left panel of Figure \ref{fig:prof} describes the typical double peak profile of a galaxy. This is the brightest source within this image and provides a good test for the quality of the reconstruction of bright sources after \texttt{MGARD} compression. The panels on the right describe the residuals of this profile between the original image and the compressed image for each error bound. The residuals are shown to be uniform and consistent with the set error bounds, noting that the compression was performed on the visibility and PSF grids and the error bounds are set on the values of these grids.

% O 12. page 7, Fig.5. How was the sigma calculated? Given that the error bounds are defined either w.r.t. the (max-min)/2 or directly, it is hard to relate the plotted
% residuals to the error bound.
%%%%% I have removed this figure as I can not improve it. Those on the right do not convey the required information.
% \begin{figure}
%     \centering
%     \includegraphics[width=\linewidth, trim=3.cm 1.cm 3.5cm 2.cm, clip=true]{profile.png}
%     \caption{ (Left) The spectral profile (relative to the RMS of the cube) of a galaxy source within the field of interest. (Right) the residuals of the profile when comparing the original cube and the cubes that have undergone \texttt{MGARD} compression.}
%     \label{fig:prof}
% \end{figure}

\subsection{Parallel I/O comparison}
%\subsection{Time comparison}
% \RD{Maybe drop this - I can't repeat it. Certainly reduce to secodn figure}
% Figure \ref{fig:benchmarking} shows the comparison between the pre-existing I/O method (labelled CASA here) and the same processing done while using ADIOS2 without engaging any parallel I/O. ADIOS2 appears to perform similarly to CASA for small amounts of data (640~MB to 6.25~GB), but improves significantly as more data (60~GB and above) is written during processing.
% \begin{figure}
%     \centering
%     \includegraphics[width=\linewidth, trim=0.cm 0.5cm 0cm 2.cm, clip=true]{benchmark_serial.png}
%     \caption{The real runtime (normalised by No. of channels per core) of the \texttt{imager} application with varying numbers of channels. CASA here refers to the standard image I/O manager and ADIOS2 refers to the ADIOS2 I/O manager. Both I/O managers have been run in serial for comparison. \RD{Needs more data points}}
%     \label{fig:benchmarking}
% \end{figure}
% Figure \ref{fig:bench_parallel} shows the comparison of the same processing as above, but using parallel-enabled ADIOS2 for writing. \AW{Add info about FitsParallel.} The same trend as before can be seen here, where ADIOS2 performs significantly better as more data is written to disk.

\begin{figure}
    \centering
    \includegraphics[width=\linewidth, trim=0.cm 0.cm 0cm 2.cm, clip=true]{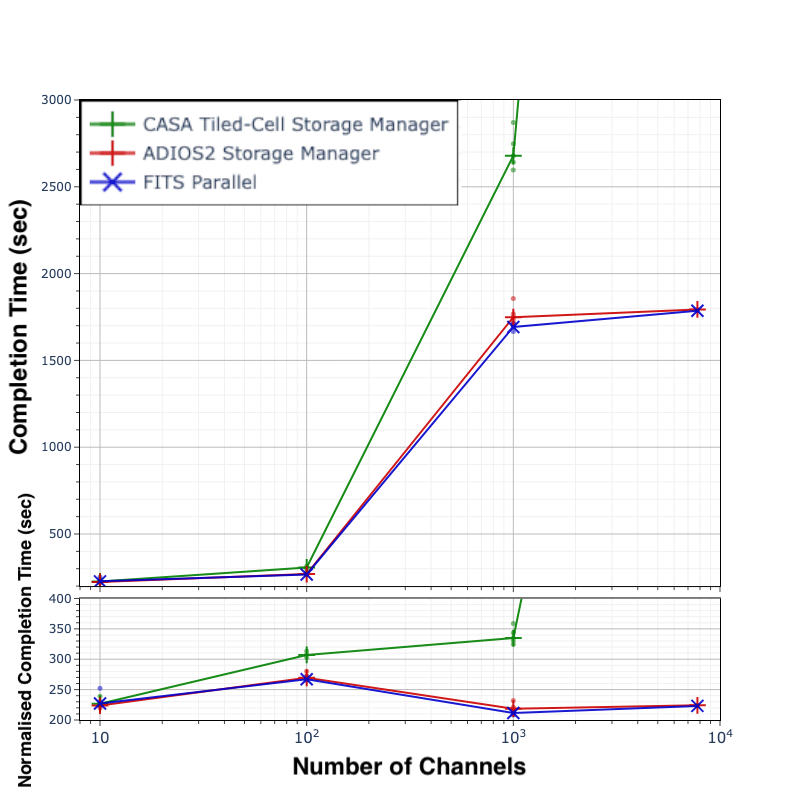}
    \caption{The real runtime (normalised by No. of channels per core) of the \texttt{imager} application 
    %in the same manner as Figure \ref{fig:benchmarking}. 
    %The real runtime  of the \texttt{imager} application with varying numbers of channels. 
    CASA here refers to the standard image I/O manager and \texttt{ADIOS2} refers to the \texttt{ADIOS2} I/O manager run in parallel, both under \texttt{CASACORE}. Fits refers to the parallel fits writer natively coded in \askapsoft{}.
    \texttt{ADIOS2} and FITS are both run in parallel via MPI whereas the CASA Tiled-Storage Manager is run in serial (due to the limitations of this storage manager). The completion time in the top panel is the real time to completion whereas the completion time in the bottom panel is normalised by the number of cores per channel. The CASA data point at 7776 channels is a lower limit (due to the processing timing out at 5 hours).}
    \label{fig:bench_parallel}
\end{figure}

Figure \ref{fig:bench_parallel} shows the comparison between the standard \texttt{CASACORE} tile manager to write the output grids,   
%and the same processing done while using ADIOS2 without engaging any parallel I/O. ADIOS2 appears to perform similarly to CASA for small amounts of data (640~MB to 6.25~GB), but improves significantly as more data (60~GB and above) is written during processing
the FitsParallel I/O libraries as used for writing large files in \askapsoft{},
and using MPI-based parallel-enabled \texttt{ADIOS2} tile manager in \texttt{CASACORE} for writing.  These are selected using the arguments in the \texttt{imager} parset file and we have taken the average of five runs to reduce timing noise on the multi-user system.
%The same trend as before can be seen here, where ADIOS2 performs significantly better as more data is written to disk.
%\RD{more discussion}
{In comparison, the single threaded CASA storage format could not compete with these parallel writing modes. 
However, we note that the most recent version of \askapsoft{} has improved the performance of this mode, but we have not yet assessed this option.}

Altering the input format of the raw data to lossy compressed format could also have a significant impact on the processing speed \citep{dodson_25} but this was not investigated, as it was out of scope. 

% 13. page 8, Fig.6. Setup of ADIOS2 image in terms of the number of cells is also of interest here. For standard FITS and standard ASKAP DINGO processing, there are 144 writers. The standard CASA image has all data in one cell. As mentioned at the beginning, serial write is more of the limitation of the single-cell CASA image format, rather than the Tiled Storage manager (which can be used to write in the distributed fashion parallelising by tiles).
%%%% at the time of the analysis (as I understand) the casa writer was not parallel - so presuably not the full implemented tiled storage manager. I will discuss this with Alex. This is now supported, as indicated in the text. 

\subsection{Comparison of compression on processing time}
\begin{figure}
    \centering
    \includegraphics[width=\linewidth, trim=0.cm 1.cm 0cm 2.cm, clip=true]{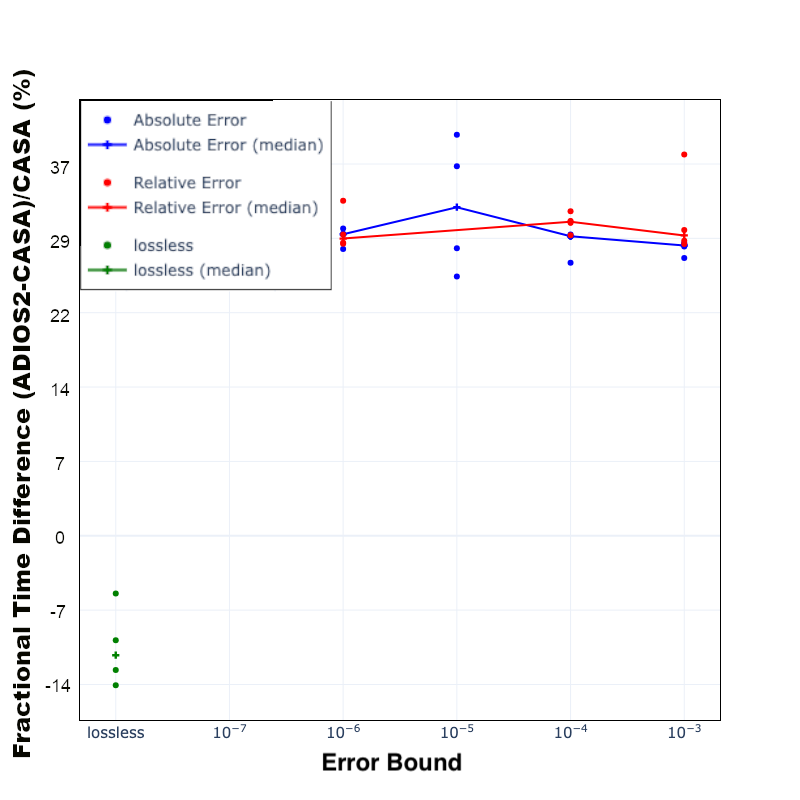}
    \caption{{The fractional processing time difference} between using the default CASA table manager and parallel threaded \texttt{ADIOS2} with different error bounds for both absolute and relative error for a 100-channel dataset, where the impact from multithreaded I/O is minimal. The median of 5 runs (points) are shown with solid lines.
    Along the x-axis we plot the error bound for different compression levels, implemented using \texttt{ADIOS2} and \texttt{MGARD} to compress the data during the writing stage of the processing.
    The fractional difference in processing time from that of the CASA manager (270\,sec) is plotted on the y-axis.
    For the lossless compression (error bound of zero) the impact is negative, meaning that the ADIOS writing is 10\% faster than the standard tilemanager, even though there is the additional compression compute required. 
    When lossy compression is used the impact is positive, meaning that the additional complication of detrending the lossy data before encoding slows the processing by about 30\%. 
    The expectation is that using the GPU code would remove or reduced this penalty. 
    \label{fig:bench_par_compress}}
\end{figure}

Figure \ref{fig:bench_par_compress} shows this comparison for the 100-channel case where the different storage managers provide comparable results, but with \texttt{MGARD} compression enabled during the writing stage of the processing. Lossless compression appears to perform significantly better than the same processing with lossy compression. 
In the current implementation, \texttt{MGARD} is run completely on CPUs, whereas the best performance comes from using the GPUs for compression in parallel. 
We plan to implement and test this compression using GPUs in the future, and expect that this will be more time-efficient. However, if the processing is I/O bound the computational costs will not be highly significant. 
    %\QG{I think the inclusion of both relative and absolute error here is misleading --- it seemed to indicate the I/O and/or compression time were different under these two settings. However, this is not the case --- during MGARD's quantisation step, the relative input error bound will be converted to absolute error bound by taking account of data's value range. Different absolute error bound can lead to varied combined I/O time as the reduced data sizes are different. Theoretically, larger error bounds shall deliver faster I/O as they reduce data more, leaving less data for writing. However, Figure 7 does not reflect this. It indicates that compression time had dominated overall I/O process for this case. I think we should add a few lines in the paper to explain and to discuss the anticipated impact on I/O using different error bounds, if the compression was implemented on GPUs.}}

% 14. page 8, Fig.7. It seems more appropriate to use the processing times ratio here rather than the difference.
%%%%%%%%%%% Agreed - but I can not easily change the figure

\subsection{Compression ratios vs distortion}
\begin{figure}
    \centering
    \includegraphics[width=\linewidth, trim=1.cm 1.cm 2.0cm 3.cm, clip=true]{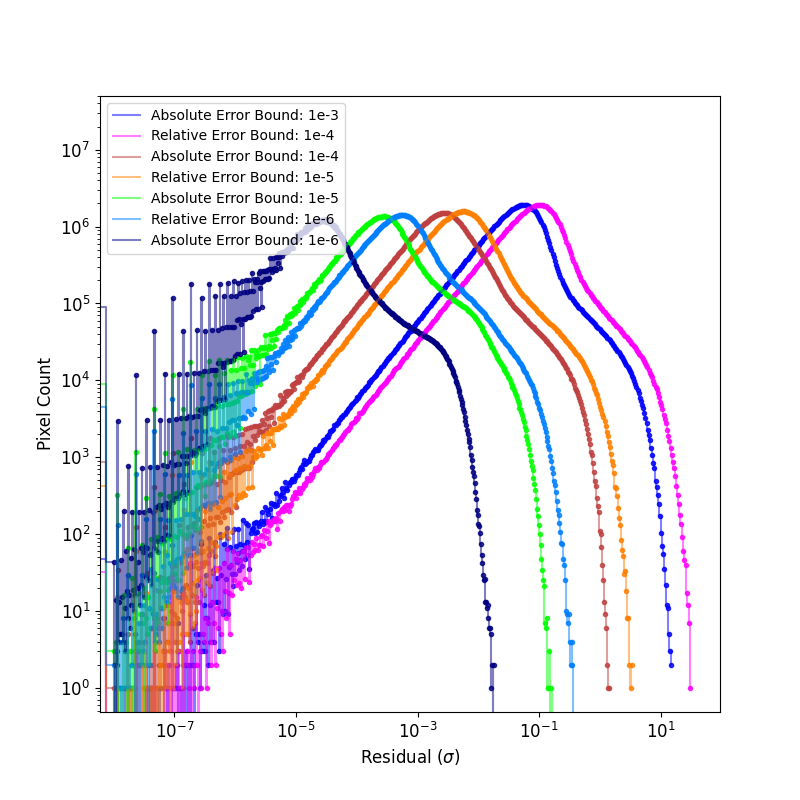}
% 16. page 9, Fig.8. Is the value on the x-axis w.r.t. the same sigma as in the earlier plots? I'm confused by the statement in the caption "the residuals are presented relative to the RMS ...".
     \caption{The distribution of residuals between the images produced with lossy compressed data and the ground truth image. The colours are as for Figure \ref{fig:imgdiff}. 
     {The residuals are normalised by the image RMS of each 100-channel cube. 
     In the first two instances (absolute error bound of $10^{-3}$ and relative error bound of $10^{-4}$) a significant number of pixel differences are greater than 1.0$\sigma$, the image RMS.}}
    \label{fig:imghistdiff}
\end{figure}

% 15. page 8, end of section 3.4. The issue with 10^{-6} absolute error bound could well be due to numerical precision of the single precision float.
%% Aggreeed added text to suggest this
The histograms of the image residuals (Figure \ref{fig:imghistdiff}) indicate the quality of the reconstructed data after imaging. \texttt{MGARD} produces a consistent residual distribution in terms of distribution shape and the maximum residuals are consistent with the specified error bounds during compression. The residuals produced from the images compressed with relative error bounds surpass one standard deviation for the $10^{-4}$ and $10^{-5}$ cases, and those for the absolute error bounds surpass $1\sigma$ for the $10^{-3}$ case.
That is, the distortions introduced by the compression, will start to be detectable against the noise levels in the images. The images shown in Figure \ref{fig:imgdiff} show that majority of these high residuals are situated in the corners of the image which are cut off or normalised out during regular imaging.

\section{Discussion and Conclusions} \label{sec:disc}

% \typeout{Compress study
% IO study
% Distortion study}

\subsection{Impact on data volumes}
The compression of the, largely empty, visibility and PSF grids is significant. The incremental compression by introducing lossy, compared to lossless, compression would allow for even smaller storage requirements. 
Compression levels of up to 20-fold do not introduce detectable errors in the reconstructed images. 
However, as there are associated risks to a multi-year observing project in using lossy compression we have decided for the DINGO project to accept the 7-fold compression we achieve with lossless compression. 
The main driver for this project was to ensure we could compress the data sufficiently to be able to complete the survey within our Pawsey storage allocation.
Lossless uv-gridding achieved this goal, and has zero risk of degradation of the data products. Thus going-forward this is the option we have adopted. 

\subsection{Impact on I/O speed}
We demonstrated the sensitivity of the processing to parallel writing in the comparison of the serial casa writer and the parallel fits or adios writer.
We note that the casa storage manager has recently been improved in \askapsoft{}, however we have not measured the performance of this new manager. 
We still favour the ADIOS library-based solution rather than a bespoke solution, firstly for the ease of code maintenance and also for the inclusion of the advanced features. 
We have found a direct impact on the total processing time with the compressed datasets; a 7-fold compression leads to a 7-fold reduction in the writing time. % to complete the task \texttt{imager}. 
% I can not find the demonstration - I don't have a suitable code base. So reduced to an obvious statement
These results will depend on the degree to which the tasks are I/O bound, the hardware of the compute, and also to degree of compression applied. 
However the reduction in processing time allows for exploration of more imaging parameters and thus a better science result.

\subsection{Impact on compute}
The computational overhead for the lossless compression is small, and is actually compensated for by {adios} I/O efficiencies. 
Lossy compression using the CPU imposes a small processing time overhead; the experience is that for GPU implementation these are even smaller \citep{Gong:2023}. %mgard_perf}. % or Gong:23?
With the resources available we were not able to explore that aspect, and we are not using the lossy compression mode in DINGO, so this was not investigated further. 

\subsection{Impact on image quality}
We confirmed that, as expected, data that has had lossless compression applied reproduced the non-compressed results perfectly. 
We have explored the image of the lossy compression for DINGO and find, for moderate compression error bounds, the impact is minimal. 
At larger compression error bounds (e.g. a relative error bound of $10^{-3}$) the image reconstruction starts to diverge. 
This we believe because of the degradation of the accuracy in the image-domain PSF when it is used for the CLEAN minor cycles. 
This is shown by the growth of the pixel error distribution shown in Figure \ref{fig:2pcorr}.
These errors build up and dominate the reconstruction. At lower error bounds the PSF is sufficiently accurate to allow for good deconvolution and image reconstruction.

\subsection{Consequences for SKA data management}
The SKA will, and have, invested a significant amount of capital into the storage, transmission and computing infrastructure for both nodes of the SKA. 
The tests described in the preceding sections have been specifically designed around the bespoke nature of deep imaging with ASKAP. 
Preliminary tests have shown that applying \texttt{MGARD} to raw SKA data (simulated) will yield similar factors of compression, the inclusion of which would simplify the implementation of storage and transmission solutions considerably for the SKA \citep{dodson_25}.
%\AW{No Longer Preliminary, update with info about visibility compression paper}.
The processing done for SKA and ASKAP to produce images is almost identical, meaning that implementing \texttt{ADIOS2} parallel I/O in the SKA pipelines would only improve the efficiency and speed of processing. 
We would recommend that this implementation is done in memory, i.e. that the xarray storage is extended to include \texttt{MGARD} compression in-situ, so that the requirements on both the physical memory and the memory bandwidth can be reduced. 

\subsection{Conclusion} \label{sec:conc}
We have successfully demonstrated the application of grid stacking observations for deep \HI{} observations using \texttt{MGARD} compression, via the \texttt{ADIOS2}/\texttt{CASACORE} framework, within standard radio astronomy software. 
The integration with \texttt{ADIOS2} provides a parallel read/writer with performance that matches the parallel FITS writer.
The combined impact will be on improved I/O performance, and thus pipeline run times, due to the reduced dataset sizes.
Comparing the images made with various compression approaches allows us to quantify the impact. 
We show that we can apply lossy compression to the data files and achieve compression ratios up to 20, using well-defined error bounds, without impacting the results. 
%The speed for the I/O remains largely unchanged. % (on the setonix file system).
Nevertheless, out of an abundance of caution, we have settled on the safer lossless compression option as that fulfils the Pawsey storage requirements for DINGO. 
In addition, the parallel reading and writing provided by ADIOS offers an additional improvement in I/O, and this is readily integrated with most common Radio Astronomy applications via the \texttt{CASACORE} libraries. 
Compression of the visibilities offers an attractive solution to the SKA I/O challenge, and we have demonstrated that the \texttt{MGARD} approach to compression can guarantee that the data is not degraded. 

\section*{Acknowledgements}
This scientific work uses data obtained from Inyarrimanha Ilgari Bundara/the Murchison Radio-astronomy Observatory. We acknowledge the Wajarri Yamaji People as the Traditional Owners and native title holders of the Observatory site. CSIRO’s ASKAP radio telescope is part of the Australia Telescope National Facility (https://ror.org/05qajvd42). Operation of ASKAP is funded by the Australian Government with support from the National Collaborative Research Infrastructure Strategy. ASKAP uses the
resources of the Pawsey Supercomputing Research Centre. Establishment of ASKAP, Inyarrimanha Ilgari Bundara, the CSIRO Murchison Radio-astronomy Observatory and the Pawsey Supercomputing Research Centre are initiatives of the Australian Government, with support from the Government of Western Australia and the Science and Industry Endowment Fund.

Part of this research was supported by the Australian Research Council Centre of Excellence for All Sky Astrophysics in 3 Dimensions (ASTRO 3D) through project number CE170100013, 
and in part by the Pawsey PACER project,
and by the U.S. Department of Energy (DOE) RAPIDS-3 SciDAC project under contract number DE-AC05-00OR22725.

Some of this work was supported by the Australian SKA Regional Centre (AusSRC), Australia’s portion of the international SKA Regional Centre Network (SRCNet), funded by the Australian Government through the Department of Industry, Science, and Resources (DISR; grant SKARC000001). AusSRC is an equal collaboration between CSIRO, Curtin University, the Pawsey Supercomputing Research Centre, and the University of Western Australia.

This research made use of Astropy, a community-developed core Python package for Astronomy \citep{Astropy-Collaboration:2013}, Matplotlib \citep{Hunter:2007}, and Numpy \citep{Harris:2020}.

\section*{Data Availability}

There is no proprietary period for ASKAP Survey Science data products, {image cubes are available from the CASDA archive but visibilities are not retained long term}. 
\New{The visibility grids for these data products are available on request to MM, and will eventually be accessible on CASDA.}
The scripts for processing and data-analysis underlying this article will be shared upon reasonable request to the corresponding author, RD.


% Don't change these lines
%\bibliography{references}
%\bibliographystyle{aasjournal}

\end{document}